\documentstyle[12pt,caption]{article}

\font\tenrm=cmr10

 1
\font\elevenrm=cmr10 scaled\magstep 1
 1

\renewenvironment{thebibliography}[1]
 { \elevenrm
   \begin{list}{\arabic{enumi}.}
    {\usecounter{enumi} \setlength{\parsep}{0pt}
     \setlength{\itemsep}{3pt} \settowidth{\labelwidth}{#1.}
     \sloppy
    }}{\end{list}}
\newcommand {\ignore}[1]{}

\newcommand{\noi}{\noindent}
\newcommand{\bc}{\begin{center}}
\newcommand{\ec}{\end{center}}

\def\ifmath#1{\relax\ifmmode #1\else $#1$\fi}
\def\3quarter{{\textstyle{3 \over 4}}}

\def\vs{\vskip}

\def\ra{\rightarrow}

\def\lf{\leaders\hbox to 1em{\hss.\hss}\hfill}

\def\21{$SU(2) \ot U(1)$}
\def\321{$SU(3) \ot SU(2) \ot U(1)$}
\def\ne{\hbox{$\nu_e$ }}
\def\nm{\hbox{$\nu_\mu$ }}
\def\nt{\hbox{$\nu_\tau$ }}
\def\ns{\hbox{$\nu_{sterile}$ }}

\def\Nt{\hbox{$N_\tau$ }}
\def\ns{\hbox{$\nu_S$ }}

\def\etal{\hbox{\it et al., }}

\def\rh{\hbox{right-handed }}

\def\gau{\hbox{gauge }}

\def\sm{\hbox{standard model }}

\def\neu{\hbox{neutrino }}
\def\sa{\hbox{such as }}

\def\neus{\hbox{neutrinos }}

\def\neusp{\hbox{neutrinos. }}

\def\eq#1{{eq. (\ref{#1})}}

\def\fig#1{{Fig. (\ref{#1})}}

\def\VEV#1{\left\langle #1\right\rangle}

\def\lsim{\raise0.3ex\hbox{$\;<$\kern-0.75em\raise-1.1ex\hbox{$\sim\;$}}}
\def\gsim{\raise0.3ex\hbox{$\;>$\kern-0.75em\raise-1.1ex\hbox{$\sim\;$}}}

\def\bel{\begin{letter}}
\def\eel{\end{letter}}
\def\beq{\begin{equation}}
\def\eeq{\end{equation}}
\def\bef{\begin{figure}}
\def\eef{\end{figure}}
\def\bet{\begin{table}}
\def\eet{\end{table}}
\def\bea{\begin{eqnarray}}
\def\ba{\begin{array}}
\def\ea{\end{array}}
\def\bi{\begin{itemize}}
\def\ei{\end{itemize}}
\def\ben{\begin{enumerate}}
\def\een{\end{enumerate}}
\def\ra{\rightarrow}
\def\ot{\otimes}

\def\eea{\end{eqnarray}}
\def\apj#1#2#3{          {\it Astrophys. J. }{\bf #1} (19#2) #3}

\def\jel#1#2#3{         {\it Journal Europhys. Lett. }{\bf #1} (19#2) #3}

\def\ib#1#2#3{           {\it ibid. }{\bf #1} (19#2) #3}
\def\nat#1#2#3{          {\it Nature }{\bf #1} (19#2) #3}
\def\nps#1#2#3{          {\it Nucl. Phys. B (Proc. Suppl.) }
                         {\bf #1} (19#2) #3}
\def\np#1#2#3{           {\it Nucl. Phys. }{\bf #1} (19#2) #3}
\def\pl#1#2#3{           {\it Phys. Lett. }{\bf #1} (19#2) #3}
\def\pr#1#2#3{           {\it Phys. Rev. }{\bf #1} (19#2) #3}

\def\prl#1#2#3{          {\it Phys. Rev. Lett. }{\bf #1} (19#2) #3}
\def\pw#1#2#3{          {\it Particle World }{\bf #1} (19#2) #3}

\def\n.c.#1#2#3{         {\it Nuovo Cim. }{\bf #1} (19#2) #3}
\def\r.n.c.#1#2#3{       {\it Riv. del Nuovo Cim. }{\bf #1} (19#2) #3}
\def\sjnp#1#2#3{         {\it Sov. J. Nucl. Phys. }{\bf #1} (19#2) #3}

\def\zetfpr#1#2#3{         {\it Z. Eksp. Teor. Fiz. Pisma. Red. }{\bf #1}
(19#2) #3}

\def\mpl#1#2#3{          {\it Mod. Phys. Lett. }{\bf #1} (19#2) #3}

\def\ppnp#1#2#3{           {\it Prog. Part. Nucl. Phys. }{\bf #1} (19#2) #3}

\def\pc{private communication}

\relax
\begin{document}
\begin{center}

{\Large \bf Recent Results in Neutrino Physics \\}
\vglue 0.5cm
{\large \bf Jos\'e W. F. Valle}
\footnote{E-mail VALLE at vm.ci.uv.es or 16444::VALLE.}
\footnote{Work supported by DGICYT and EEC grants PB92-0084
and CHRX-CT93-0132.}\\
\vglue 0.5cm
{\sl Instituto de F\'{\i}sica Corpuscular - C.S.I.C.,
Departament de F\'{\i}sica Te\`orica, Universitat de Val\`encia\\
46100 Burjassot, Val\`encia, SPAIN         }\\
\vskip 1cm
{\bf ABSTRACT\\}
\end{center}
 \tenrm\baselineskip=15pt
 \noindent
Present limits on \neu masses are reviewed,
along with the positive cosmological and astrophysical
hints from dark matter, solar and atmospheric neutrino
observations. If all these hints are due to \neu
physics, either \neus are closely degenerate, with
a mass of about 2 eV, leading to neutrinoless $\beta\beta$
decay rate observable in the next round of experiments, or
else a light sterile \neu exists in nature.
In either case the simplest seesaw scheme would
be ruled out. However one may consistently implement
the quasidegenerate \neu scenario in extended seesaw
models, eg based on SO(10).
The light sterile \neu possibility can be implemented in
schemes with radiative mass generation, leading to
the possibility of enhanced lepton flavour violating
processes as well as \neu oscillations observable at
accelerators.
Finally I discuss an direct, but striking, possible
manifestation of \neu masses in the symmetry breaking
sector of the electroweak theory: the invisibly decaying
higgs boson. I describe how LEP data can be used to
provide model independent limits on the higgs particle
and also discuss the prospects for probing the
associated physics at higher energies.

\section{Preliminaries}

No solid theoretical principle prevents \neus from having mass.
Moreover, from the point of view of theory, it is rather
mysterious that \neus seem to be so special when compared
with the other fundamental fermions. Many attractive
extensions of the \sm require \neus to be massive \cite{fae}.
This is the case, for example, in SO(10) or left right
symmetric theories, where the presence of \rh neutrinos
is required in order to realize the extra symmetry. On the
other hand there is, in these theories, a natural mechanism to
understand the relative smallness of \neu masses \cite{GRS}.
In this case lepton number is part of the \gau symmetry and
its feeble violation is related to the observed smallness
of \neu masses and to the V-A nature of the weak interactions
\cite{LR}.

This is by no means the only way to \neu masses.
Indeed, it has been realized in the early days
that lepton number may be a spontaneously broken
global symmetry \cite{CMP0}. Since then there have
been many other attractive suggestions of how to realize
this idea in realistic scenarios \cite{fae}. In this
case, quite naturally, the observed smallness of \neu
masses does not require any large mass scale.
The extra particles required to generate the \neu
masses have masses at scales accessible to present
experiments. Such a low scale for lepton number breaking
could have important implications not only in astrophysics
and cosmology (e.g. electroweak baryogenesis) but also in
particle physics as we will discuss below.

Whichever way one adopts, present theory is not capable,
from general principles, of predicting the scale of \neu
masses any better than it can fix the masses of the other
quarks and charged leptons, say the muon. One should at this
point turn to experiment.

There are several limits on \neu masses that follow
from observation. The laboratory bounds may be
summarized as \cite{PDG92}
\beq
\label{1}
m_{\nu_e} 	\lsim 10 \: \rm{eV}, \:\:\:\:\:
m_{\nu_\mu}	\lsim 270 \: \rm{keV}, \:\:\:\:\:
m_{\nu_\tau}	\lsim 31  \: \rm{MeV}
\eeq
These limits follow purely from kinematics
and have therefore the great advantage that they
are the most model-independent of the \neu mass
limits. The experimental status of the limits on
the \ne mass have been extensively discussed
here \cite{Erice}.
Note that the limit on the \nt mass may
be substantially improved at a tau factory \cite{jj}.
In addition, there are limits
on neutrino masses that follow from the nonobservation of
neutrino oscillations. I address you to ref.
\cite{granadaosc} for a detailed discussion
and compilation. As opposed to the limits in \eq{1}
\neu oscillation limits are correlated ones, involving
\neu mass differences versus mixing. Thus they rely on
the additional assumption, although quite natural in
\gau theories, that massive \neus do mix.

Apart from the above limits, there is an important
one derived from the non-observation of the
${\beta \beta}_{0\nu}$ nuclear decay process i.e.
the process by which nucleus $(A,Z-2)$ decays to
$(A,Z) + 2 \ e^-$.
This lepton number violating process would arise
via \neu exchange and, although highly favoured by phase space
over the usual $2\nu$ mode, it proceeds only if the virtual neutrino
is a Majorana particle. The decay amplitude is
proportional to
\beq
\VEV{m} = \sum_{\alpha} {K_{e \alpha}}^2 m_{\alpha}
\label{AVERAGE}
\eeq
where $\alpha$ runs over the light neutrinos.
The non-observation of ${\beta \beta}_{0\nu}$
in $^{76} \rm{Ge}$ and other nuclei leads to the limit \cite{Avignone}
\beq
\label{bb}
\VEV{m} \lsim 1 - 2 \ eV
\eeq
depending on nuclear matrix elements \cite{haxton_granada}.
Even better sensitivity is expected from the upcoming
enriched germanium experiments \cite{Avignone}.
Although rather stringent, the limit in \eq{bb}
is rather model-dependent, and does not apply
when total lepton number is an unbroken symmetry,
as is the case for Dirac \neusp Even if all \neus
are Majorana particles, $\VEV{m}$ may differ substantially
from the true neutrino masses $m_\alpha$ relevant for kinematical
studies, since in \eq{AVERAGE} the contributions of
different neutrino types may interfere destructively,
similarly to what happens in the simplest Dirac \neu case,
where the lepton number symmetry enforces that
$\VEV{m}$ automatically vanishes \cite{QDN}.

The ${\beta \beta}_{0\nu}$ decay process may
also be engendered through the exchange of scalar
bosons, raising the question of which relationship the
${\beta \beta}_{0\nu}$ decay process bears
with the \neu mass.
A simple but essentially rigorous proof shows that,
in a gauge theory, whatever the origin of ${\beta \beta}_{0\nu}$
is, it requires \neus to be Majorana particles, as illustrated in \fig{box}.
\bef
\vspace{4.5cm}
\caption{${\beta \beta}_{0\nu}$ decay and Majorana neutrinos.}
\label{box}
\eef
Indeed, any generic "black box" mechanism inducing
neutrinoless double beta decay can be closed, by W exchange,
so as to produce a diagram generating a nonzero Majorana
neutrino mass, so the relevant neutrino will,
at some level, be a Majorana particle \cite{BOX}.

Gauge theories may lead to new varieties of
neutrinoless double beta decay involving the
emission of light superweakly interacting spin
zero particles \cite{GGN}. One of these, called
majoron, is the goldstone boson associated
to the spontaneous violation of a global
lepton number symmetry \cite{CMP0}
\footnote{A related light scalar
boson $\rho$ should also be emitted.}
\beq
(A,Z-2) \rightarrow (A,Z) + 2 \ e^- + J \:.
\eeq
The emission of such light scalars would only be detected
through their effect on the $\beta$ spectrum.

The simplest model with sizeable majoron
emission in $\beta\beta$ decays involving an
isotriplet majoron \cite{GR} leads to a new
invisible decay mode for the neutral \gau
boson with the emission of light scalars,
\beq
Z \ra \rho + J,
\label{RHOJ}
\eeq
now ruled out by LEP measurements of the
invisible Z width \cite{LEP1}.

However it has been recently shown that a sizeable
majoron-neutrino coupling leading to observable
emission rates in neutrinoless double beta decay
can be reconciled with the LEP results in models
where the majoron is an isosinglet and lepton number
is broken at a very low scale \cite{ZU}. An alternative
possibility was discussed in \cite{Burgess93}.
Recently there have been negative searches for
the majoron emitting neutrinoless double beta decay
by the Irvine and Heidelberg-Moscow groups
which lead to a limit on the majoron-neutrino
coupling of about $10^{-4}$ \cite{klapdor_wein}.

In addition to laboratory limits, there is a cosmological
bound that follows from avoiding the overabundance of
relic neutrinos \cite{KT}
\beq
\sum_i m_{\nu_i} \lsim 50 \: \rm{eV}
\label{rho1}
\eeq
This limit is also model-dependent, as it only holds
if \neus are stable on cosmological time scales.
There are many models where neutrinos decay into
a lighter \neu plus a majoron \cite{fae},
\beq
\nu_\tau \ra \nu_\mu + J \:\: .
\label{NUJ}
\eeq
Lifetime estimates in seesaw type majoron models have
been discussed in ref. \cite{V}. Here I borrow the estimate
of the model of ref. \cite{ROMA}, given by curve C in \fig{ntdecay}.
The solid line gives the lifetime required in order
to suppress the relic \nt contribution.
The dashed line ensures that the universe has become
matter-dominated by a redshift of 1000 at the latest
so that fluctuations have grown by the same factor by
today  \cite{ST}
\footnote{However, this lifetime limit
is less reliable than the one derived from the critical
density, as there is not yet an established theory for the
formation of structure in the universe.}.
Comparing curve C with the solid and dashed lines
one sees that the theoretical lifetimes can be shorter
than required. Moreover, since these decays are
$invisible$, they are consistent with all
astrophysical observations.
Recently Steigman and
collaborators have argued that many values of the \nt mass
can be excluded by cosmological big-bang nucleosynthesis,
even when it decays \cite{BBNUTAU}. This, however, still
leaves open a wide region of theoretically interesting
\nt lifetime-mass values.
\bef
\vspace{8.4cm}
\caption{
Estimated \nt lifetime versus observational limits.
}
\label{ntdecay}
\eef
It follows than that any effort to improve present
\neu mass limits is worthwhile. These include searches
for distortions in the energy distribution of the electrons
and muons coming from decays \sa
$\pi, K \ra e \nu$, $\pi, K \ra \mu \nu$, as
well as kinks in nuclear $\beta$ decays \cite{Deutsch}.

\section{Positive Hints for Neutrino Mass}

\noi
In addition to the {\sl limits} described in the
previous section, observation also provides us with
some positive {\sl hints} for neutrino masses.
These follow from cosmological, astrophysical
and laboratory observations which I now discuss.

Recent observations of cosmic background temperature
anisotropies on large scales by the COBE  satellite,
when combined with smaller scale observations
(cluster-cluster correlations) indicate the need for
the existence of a hot {\sl dark matter} component,
contributing about 30\% to the total mass density,
i.e. $\Omega_{HDM} \sim 0.3$ \cite{cobe,cobe2}.
For this the most attractive particle candidate is a
massive neutrino, \sa as a \nt of a few eV mass.
This suggests the possibility of having
observable \ne to \nt or \nm to \nt oscillations in the
laboratory. The next generation of experiments CHORUS
and NOMAD at CERN, and the P803 experiment proposed at
Fermilab will probe this possibility \cite{chorus}.

Second, the {\sl solar neutrino data} collected up to now by
the two high-energy experiments Homestake and Kamiokande,
as well as by the low-energy data on pp neutrinos from
the GALLEX and SAGE experiments still pose a persisting
puzzle \cite{Davis,granadasol}. Comparing the full data
of GALLEX including their most recent ones, with the
Kamiokande data, one can obtain the allowed one sigma
region for $^7 $ Be and $^8$ Be fluxes as the intersection
of the region to the left of line labelled 91 with
the region labelled KAMIOKA.
The lines are normalized with respect to the reference
solar model of Bahcall and collaborators. Including
the Homestake data of course only aggravates the
discrepancy \cite{Smirnov_wein}, as can be seen
from the fig xx.
\bef
\vspace{9cm}
\caption{
Allowed one sigma bands for $^7 $ Be and $^8$ Be fluxes
from all solar neutrino data
}
\label{solardata}
\eef
Thus the solar \neu problem seems really a problem.
The simplest astrophysical solutions are highly disfavored
if {\sl all} data are taken simultaneously, leading to the
need of new physics in the \neu sector \cite{NEEDNEWPHYSICS}.
The most attractive way to account for the data
is to assume the existence of \neu conversions
involving very small \neu masses $\sim 10^{-3}$ eV
\cite{MSW}. The region of parameters allowed by
present experiments is illustrated in \fig{msw}
\cite{Hata} (for similar analyses, see ref.
\cite{MSWPLOT}). Note that the fits favour the
non-adiabatic over the large mixing solution,
due mostly to the larger reduction of
the $^7 $ Be flux found in the former.
\bef
\vspace{9.5cm}
\caption{Region of solar \neu oscillation parameters
allowed by experiment}
\label{msw}
\eef

Finally, there are hints for \neu masses from studies involving
{\sl atmospheric neutrinos}. Although the predicted absolute
fluxes of \neus produced by cosmic-ray interactions in the
atmosphere are uncertain at the 20 \% level, their
ratios are expected to be accurate to within
5 \% \cite{atmsasso}.
An apparent decrease in the expected flux of atmospheric
$\nu_\mu$'s relative to $\nu_e$'s arising from the decays
of $\pi$'s and $K$'s produced in the atmosphere, and from
the secondary muon decays has been observed in three
underground experiments, Kamiokande, IMB and possibly
Soudan2 \cite{atm}. This atmospheric neutrino deficit
can be ascribed to \neu oscillations.
Combining these experimental results with observations
of upward going muons made by Kamiokande, IMB and Baksan,
and with the negative Frejus and NUSEX results \cite{up}
leads to the following range of neutrino oscillation
parameters \cite{atmsasso}
\beq
\label{atm0}
\Delta m^2_{\mu \tau} \approx 0.005 \: - \: 0.5\ \rm{eV}^2,\
\sin^22\theta_{\mu \tau} \approx 0.5
\eeq
These recent analyses severely constrain the oscillation
parameters, apparently excluding oscillations of \nm to \nt
with maximal mixing, as expected in some theoretical models.
However, the underlying uncertainties are still so large that it is
unsafe to rule out maximal mixing with a high degree of confidence.
Similar analyses have also been performed for the case of
\nm to \ns as well as \nm to \ne channels, where matter
effects play an important role \cite{lipari}.

Taken at face value, the above astrophysical and cosmological
observations suggest an interesting theoretical puzzle, if one
insists in accounting for all three observations on solar,
dark matter and atmospheric \neus within a consistent theory.
Indeed, it is difficult to reconcile these three observations
simultaneously in the framework of the simplest seesaw model
with just the three known \neus. The only possibility is if
all three \neus are closely degenerate \cite{caldwell}.

We now turn to model building. Can we reconcile the present
hints from astrophysics and cosmology in the freamweork of a
consistent elementary particle physics model?

It is known that the general seesaw models have
two independent terms giving rise to the light \neu masses.
The first is an effective triplet vacuum expectation value
\cite{2227} which is expected to be small in left-right
symmetric models \cite{LR}. Based on this fact one can
in fact construct extended seesaw models where the main
2 eV or so contribution to the light \neu masses is universal,
due to a suitable horizontal symmetry, while the splittings
between \ne and \nm explain the solar \neu deficit and that
between \nm and \nt explain the atmospheric \neu anomaly \cite{DEG}.

The alternative way to fit all the data is to add a
fourth \neu species which, from the LEP data on the
invisible Z width, we know must be of the sterile type,
call it \ns. Two basic schemes have been suggested in
which the \ns either lies at the dark matter scale
\cite{DARK92} or, alternatively, at the solar \neu
scale \cite{DARK92B}.
In the first case the atmospheric
\neu puzzle is explained by \nm to \ns oscillations,
while in the second it is explained by \nm to \nt
oscillations. Correspondingly, the deficit of
solar \neus is explained in the first case
by \ne to \nt oscillations, while in the second
it is explained by \ne to \ns oscillations. In both
cases it is possible to fit all observations together.
However, in the first case there is a clash with the
bounds from big-bang nucleosynthesis while, in the
latter case of where \ns is at the MSW scale these
limits can be used to single out the nonadiabatic
solution uniquely. Note however that, since the
mixing angle characterizing the \nm to \nt
oscillations is nearly maximal, the second
solution is in apparent conflict with \eq{atm0}.
Another theoretical possibility is that all active
\neus are very light but the sterile \neu \ns is
the single \neu responsible for the dark matter
\cite{DARK92D}.

In short, \neu masses, besides being suggested
by theory, seem to be required to fit present
astrophysical and cosmological observations.
The solid curves in fig. 5 show the regions of
\ne to \nm and \nm to \nt oscillation parameters
that are excluded by present accelerator and
reactor experiments.
The next generation of accelerator
experiments at CERN may test for the existence
of \neu oscillations involving the \nt. This
is indicated by the dot-dashed line in figure 5.
Finally, the regions suggested by present solar
and atmospheric \neu data are sketched, for comparison.
Regions A and B are the allowed MSW solutions for solar
\neus while the unlabeled regions are for atmospheric \neus.
Similar plots can be made for the case of sterile \neus.
\bef
\vspace{9cm}
\caption{
Neutrino oscillation parameters, present hints,
limits, and future experimental sensitivities.
See text.}
\label{osci}
\eef
Further progress will be achievable at the upcoming
long baseline experiments planned at BNL, Soudan,
Icarus and Kamiokande (dashed lines). Underground experiments
should also help to clarify whether or not solar
\neu conversions exist and also search for
neutrinoless double beta decay with sensitivity
enough to probe the quasidegenerate \neu scenario
outlined above.

In addition to 	\neu oscillations,
there are many other lepton flavour violating
processes whose existence would be related
to neutrino masses and neutrino properties
beyond the standard model. These include
	$\mu \ra e \gamma$,
	$\mu \ra 3 e $,
	$\mu \ra e$ conversion in nuclei,
	$\tau \ra e \pi^0$,
	$\tau \ra e \gamma$,
as well as two-body decays with the emission
of a superweakly interacting majoron,
e.g. $\mu \ra e + J$ and $\tau \ra e,\mu + J$.
The underlying physics may also be probed
at the high energies accessible at LEP, through related
Z decay processes. e.g. $Z \ra \Nt \nt$ or
$Z \ra \chi \tau$, where \Nt denotes a neutral heavy
lepton, while $\chi$ denotes the lightest chargino.
All of these processes may occur at levels consistent with
present or planned experimental sensitivities, without
violating any experimental data. For recent discussions
see ref. \cite{fae,lfv94}.

\section{Invisible Higgs Decays}

We now turn to a much less usual and less
direct, but striking, possible manifestation of
\neu masses in the symmetry breaking sector of the
electroweak theory.

In many models \cite{JoshipuraValle92} \neu masses
are induced from the spontaneous violation of a global
$U(1)$ lepton number symmetry by an \21 singlet vacuum
expectation value $\VEV{\sigma}$, in such a way that
$m_\nu \to 0$ as $\VEV{\sigma} \ra 0$.
In contrast with the more usual seesaw majoron model
\cite{CMP0}, a low scale for the lepton number violation,
close to the electroweak scale, is {\sl preferred} in
these models, since it is required in order to obtain
small neutrino masses \cite{JoshipuraValle92}
\footnote{Another example is provided by the
RPSUSY models \cite{HJJ}.}.
Another cosmological motivation for low-scale
majoron models has been given in ref. \cite{Goran92}.

In these models, although the majoron has very tiny couplings to
matter, it can have significant couplings to the
Higgs bosons.
This implies that the Higgs boson may decay
with a substantial branching ratio into the
invisible mode \cite{JoshipuraValle92,Joshi92}
\begin{equation}
h \rightarrow J\;+\;J
\label{JJ}
\end{equation}
where $J$ denotes the majoron. The presence of
this invisible Higgs decay channel can affect
the corresponding Higgs mass bounds in an
important way, as well as lead to novel
search strategies at higher energies.

The production and subsequent decay of any Higgs boson
which may decay visibly or invisibly involves three independent
parameters: the Higgs boson mass $M_H$, its coupling
strength to the Z, normalized by that of the \sm, call
this factor $\epsilon^2$, and the invisible Higgs boson
decay branching ratio.

The results published by the LEP experiments on the
searches for various exotic channels can be used
in order to determine the regions in parameter space
that are ruled out already. The procedure was described
in \cite{alfonso,dproy}. Basically it combines the results
of the standard model Higgs boson searches with those one
can obtain for the invisible decay.
For each value of the Higgs mass, the lower bound on
$\epsilon^2$ can be calculated as a function of the
branching ratio $BR(H \rightarrow $ visible), both this
way as well as through the \sm Higgs search analyses
techniques. The weakest of such bounds for
$BR(H \rightarrow $ visible) in the range
between 0 and 1, provides the absolute bound on $\epsilon^2$.
This procedure can be repeated for each value of $M_H$, thus
providing an an exclusion contour in the plane $\epsilon^2$
vs. $M_H$, shown in \fig{alfonso2}, taken from ref. \cite{alfonso}.
The region in $\epsilon^2$ vs. $M_H$ that is already excluded by the
present LEP analyses holds {\sl irrespective of the mode of Higgs decay},
visible or invisible.
\bef
\vspace{7cm}
\caption{Region in the $\epsilon^2$ vs. $m_H$ that can be
excluded by the present LEP1 analyses, independent of the
mode of Higgs decay, visible or invisible (solid curve).
Also shown are the LEP2 extrapolations (dashed).}
\label{alfonso2}
\eef
Finally, one can also determine the additional
range of parameters that can be covered by LEP2
for a total integrated luminosity of 500 pb$^{-1}$
and centre-of-mass energies of 175 GeV and 190 GeV.
This is shown as the dashed and dotted curves in
\fig{alfonso2}.

The possibility of invisible Higgs decay
is also very interesting from the point of
view of a linear $e^+ e^-$ collider at higher
energy \cite{EE500}. Heavier, intermediate-mass,
invisibly decaying Higgs bosons can also be searched at high energy
hadron supercolliders such as LHC/SSC \cite{granada}.
The limits from LEP discussed above should
serve as useful guidance for such future searches.

\section{Conclusion}

\noi
Present cosmological and astrophysical observations,
as well as theory, suggest that neutrinos may be massive.
Neutrino masses might even affect the electroweak
symmetry breaking sector in a very important way.

Existing data do not preclude neutrinos from being responsible
for a wide variety of measurable implications at the laboratory.
These new phenomena would cover an impressive region of energy,
from $\beta$ and double $\beta$ decays, to neutrino oscillations,
to rare processes with lepton flavour violation, up to LEP
energies.
The next generation of \neu oscillation searches
sensitive to \nt as dark matter (CHORUS/NOMAD/P803),
$e^+ e^-$ collisions from ARGUS/CLEO to tau-charm and
B factories, as well as the experiments at LEP and the
future LHC could all be sensitive to \neu properties!

It is therefore worthwhile to keep
pushing the underground experiments, for possible
confirmation of \neu masses. The neutrinoless
$\beta\beta$ decay searches with enriched germanium
could test the quasidegenerate neutrino scenario for
the joint explanation of hot dark matter and
solar and atmospheric \neu anomalies.
Further data from low energy pp neutrinos
as well as from Superkamiokande, Borexino, and
Sudbury will shed further light on the neutrino
sector. The same can be said of the ongoing
studies with atmospheric \neusp

Similarly, a new generation of experiments capable
of more accurately measuring the cosmological
temperature anisotropies at smaller angular scales than
COBE, would be good probes of different models of
structure formation, and presumably shed further
light on the need for hot \neu dark matter.
All such endeavours should be gratifying!

\vs .3cm

{\large Acknowledgements}
\vs .3cm

I thank Fernando de Campos for reading the
manuscript and helping compose it. I also
thank the organizers for the kind invitation
and friendly organization.
\noi
\vglue 0.6cm
{\large \bf References}
\bibliographystyle{ansrt}

\end{document}